\documentclass[12pt,preprint]{aastex}

\shorttitle{Unveiling the NLSy1 Nature of Mrk573.}
\shortauthors{Ramos Almeida et al.}

\begin{document}

\title{Unveiling the Narrow-Line Seyfert 1 Nature of Mrk 573 using near-Infrared spectroscopy.}

\author{C. Ramos Almeida\altaffilmark{1}, A. M. P\'{e}rez Garc\'\i a\altaffilmark{1}, 
J. A. Acosta-Pulido\altaffilmark{1}, and O. G\'onzalez-Mart\'\i n\altaffilmark{2}}

\altaffiltext{1}{Instituto de Astrof\'\i sica de Canarias (IAC), 
              C/V\'\i a L\'{a}ctea, s/n, E-38205, La Laguna, Tenerife, Spain.
cra@iac.es, apg@iac.es, jap@iac.es}
\altaffiltext{2}{Instituto de Astrof\'\i sica de Andaluc\'\i a, CSIC, Po.Box 3004, E-18080 
              Granada, Spain.
omaira@iaa.es}

\begin{abstract}

In this letter we present clear evidence that Mrk~573 is an obscured 
Narrow-Line Seyfert 1 (NLSy1) and not an archetypal Seyfert 2, 
as it has been classified until now. Only three galaxies have been 
proposed as members of this class, prior to this work. 
Here we report near-infrared spectroscopic data taken with  LIRIS 
on the 4.2 m William Herschel Telescope (WHT).
Our high quality near-infrared nuclear spectrum in the 0.88-1.35~\micron\ wavelength range 
shows the permitted O~I $\lambda\lambda$1.128,1.317 narrow lines, the Fe II 9200-\AA~lines and 
the Fe II 1-\micron~lines, together with a relatively broad component ($\sim$~1700 km~s$^{-1}$) of Pa$\beta$. 
These features can originate only in an optically thick high density region, similar to those observed 
in the Broad Line Region (BLR). 

\end{abstract}


\section{Introduction}
 
The NLSy1 galaxies \citep{Osterbrock85,Pogge00} share the optical properties of Seyfert 1 nuclei, 
except that they show narrower Balmer lines (FWHM $<$ 2000 km~s$^{-1}$) 
and strong optical Fe II emission. 
In the X-rays, these objects are also peculiar because of their soft X-ray excess and 
rapid X-ray variability \citep{Boller96,Véron01}, together with a very steep photon spectral index 
\citep{Leighly99}. The most accepted explanation for these extreme X-ray properties are 
high Eddington ratios (L/L$_{Edd}~\sim$ 1-10) and lower black hole masses relative to other 
Seyfert nuclei (M$_{BH}~\sim~10^{6-7} M_{\sun}$) \citep{Mathur01,Boroson02,Grupe04}. 
In addition, \citet{Nagar02} and \citet{Dewangan05} suggested the existence of the Type 2 counterparts 
of the NLSy1, i.e., obscured NLSy1, which appear as Seyfert 2 galaxies but containing 
a Seyfert 1 engine characterized by a high accretion rate or a small BH mass. 
\citet{Dewangan05} proposed NGC~7582, NGC~5506, and NGC~7314 as obscured NLSy1,
based on their extreme X-ray variability and Compton-thick nature. NGC~5506 was also confirmed 
as a NLSy1 in the near-infrared by \citet{Nagar02}.
This type of objects would be missclassified using only
the visible range. They can be distinguished through low extinction wavelength
regimes, for example the near-infrared.

The galaxy Mrk~573 has been extensively studied by many authors. Its active nucleus is 
hosted in a (R)SAB(rs)0+ galaxy\footnote{The redshift of the 
galaxy has been measured to be  z=0.017 \citep{Ruiz05}, which implies a
physical scale of 333 pc~arcsec$^{-1}$, using H$_{0}$ = 75 km s$^{-1}$ Mpc$^{-1}$.} and 
has been classified until now as a classical Type 2 nucleus from previous optical 
\citep{Tsvetanov92,Erkens97,Mullaney08} and infrared spectroscopic data 
\citep{Veilleux97,Riffel06}. 
Spectropolarimetric observations \citep{Nagao04} reveal a broad component of 
the H$\alpha$ profile ($\sim$~3000 km~s$^{-1}$), together with the optical Fe II multiplet.  
Prominent forbidden high-ionization emission lines have also been detected in the optical 
spectrum of this galaxy, as for example [Fe VII]$\lambda\,0.609$ and [Fe X]$\lambda\,0.637$ 
\citep{Erkens97,Mullaney08}.
In the X-ray regime this galaxy presents a soft X-ray excess (L$_{soft}^{obs}$=1.6x10$^{41}$ erg~s$^{-1}$) 
and a steep photon spectral index ($\Gamma$=2.7$\pm$0.4) \citep{Guainazzi05}. 
It also appears as a Compton-thick object 
($\mathrm{N_{H}} > 1.6\times 10^{24}~\mathrm{cm}^{-2}$) 
and shows a large EW of the K$_\alpha$ fluorescent iron line \citep{Guainazzi05}. 

In this letter we report on new LIRIS near-infrared spectroscopy of Mrk~573, 
that, together with the X-ray properties of this galaxy, 
unambiguously identifies it as an obscured NLSy1. 
This re-classification is based on the detection of the permitted Fe II 9200-\AA~lines and 
the Fe II 1-\micron~lines, together with O I $\lambda\lambda$1.128,1.317, 
and a relatively broad component of the Pa$\beta$ line. 
We also report the first detection of a conspicuous line at 0.962 \micron, that 
tentatively could be identified either as O I and/or He I transitions.

\section{Observations and data reduction}

	High quality near-infrared long-slit spectra in the range 
0.8-2.4 \micron\ were obtained on the night of 
2006 July 17, using the near-infrared camera/spectrometer LIRIS 
\citep{Manchado04}, on the
4.2 m William Herschel Telescope (WHT). The spatial scale is
0.25\arcsec~pixel$^{-1}$, and the slit width 0.75\arcsec~(allowing a
spectral resolution in the range $\sim$~600-400 km~s$^{-1}$). The slit was centered 
on the galaxy nucleus and it was oriented along P.A.=122$^{o}$, coincident with the orientation 
of the ionization cone observed by \citet{Pogge95}. 
Weather conditions were good, with the seeing varying between 0.8\arcsec and 0.9\arcsec. 
Observations were performed following an ABBA telescope-nodding pattern. 
A total exposure time of 4000~s was taken, divided 
into individual frames of 500~s each. 
The data were reduced following standard procedures for near-infrared spectroscopy, 
using the {\it lirisdr} dedicated software within the 
IRAF\footnote{IRAF is distributed by the National Optical Astronomy Observatory, which is 
operated by the Association of Universities for the Research in Astronomy, Inc., under 
cooperative agreement with the National Science
Foundation (http://iraf.noao.edu/).} enviroment. 
For a detailed description of the reduction process, see \citet{Ramos06}.
The nearby A0 HD~2193 star was observed with the same configuration and similar airmass 
to the galaxy to perform the telluric correction  and the 
flux calibration. A modified version of {\it Xtellcor} \citep{Vacca03} was used in this step.
Note that the absolute flux calibration is intended to be an approximation since 
the spectrum of the comparison star is probably subject to centering and tracking 
errors, as well as slit losses.

\section{Results}

\begin{figure*}
\centering
\includegraphics[width=9cm,angle=90]{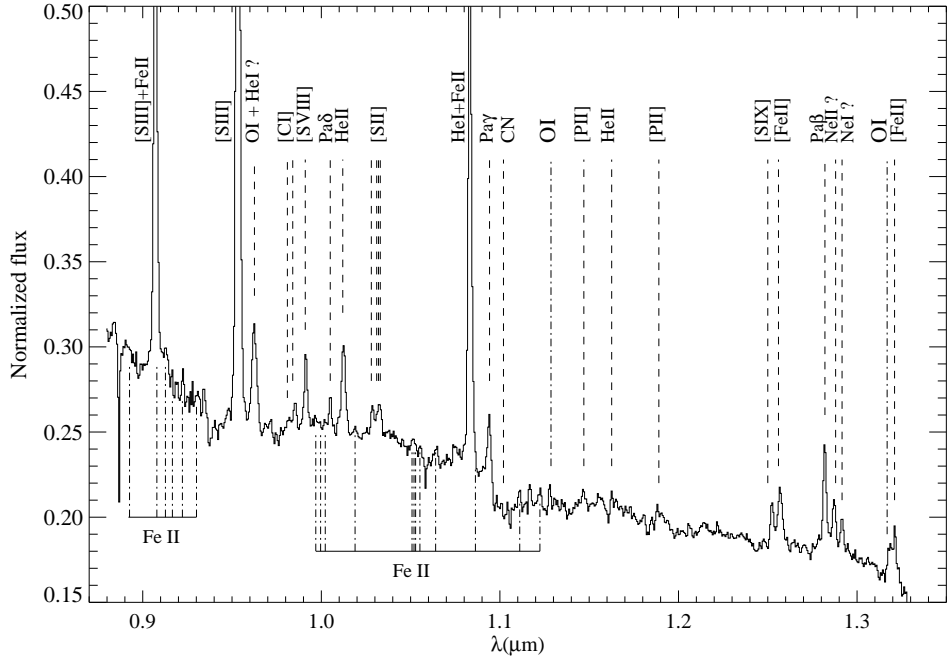}
\caption{\footnotesize{Normalized spectrum of the nuclear region of Mrk~573 in the ZJ range. 
Detections of the permitted Fe II 9200-\AA~lines and 
the Fe II 1-\micron~lines, together with the O I $\lambda\lambda$1.128,1.317 are clear. Note the broad base of the
Pa$\beta$ profile.}
\label{spectrum}}
\end{figure*}

	The final nuclear spectrum, extracted with a 1.5\arcsec~aperture centered 
on the maximum of the galaxy profile, is shown in Fig.~\ref{spectrum}. 
Most detected lines are marked in the figure, and their fluxes and FWHM 
(measured by fitting a Gaussian profile using the Starlink program DIPSO) are reported 
on Table \ref{fluxes}. In general terms, the spectrum is typical of a Seyfert 2
galaxy, showing very high ionization lines such as [S\,VIII]$\lambda 0.991$ and 
[S\,IX]$\lambda 1.252$. One can also note the presence of the 1.1 \micron~CN band, first reported
for this galaxy by \citet{Riffel07}, that shows the existence of young/intermediate stars in the nuclear
region.
The main focus of this letter is on the detection of the permitted O~I and 
Fe~II lines well above the noise level and of the broad wings of the Pa$\beta$ profile.

The O~I$\lambda\lambda$1.128,1.317 permitted transitions are produced by Ly$\beta$ pumping in a 
Bowen fluorescence mechanism \citep{Grandi80}, and they are only produced in a high density 
optically-thick gas. 
Thus, these lines are seen only in Seyfert 1 or NLSy1 nuclei, but never in canonical Seyfert 2 galaxies.

\begin{figure}
\centering
\includegraphics[width=7cm,height=8cm,angle=90]{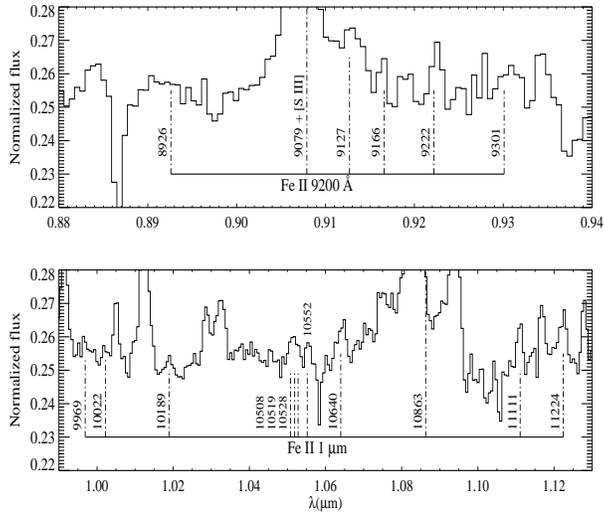}
\caption{\footnotesize{Detailed view of the nuclear spectrum of Mrk~573 around the 
permitted Fe II 9200-\AA~lines and the Fe II 1-\micron~lines.}
\label{iron}}
\end{figure}

In Fig.~\ref{iron} all the detected Fe II transitions are labelled 
for both the Fe II 9200-\AA~lines (0.893, 0.908, 0.913, 0.917, 0.922, 0.930 \micron), 
and the Fe II 1-\micron~lines (0.997, 0.999, 1.002, 1.019, 1.051, 1.052, 1.053, 1.055, 1.064, 1.086, 
1.111, 1.122 \micron).
Very few Seyferts with near-infrared detection of permitted Fe II lines have been reported until now in the
literature \citep{Rudy00,Rudy01,Nagar02,Rodriguez02,Rodriguez05}.	
The detection of Fe II lines provides crucial information on the 
ionization mechanism of the gas, since they are thought to be pumped by Ly$\alpha$ fluorescence 
\citep{Johansson84,Sigut03} and  they can be formed only in high density clouds  or in the 
optically-thick BLR.

For Mrk~573, we estimate the line width of O~I$\lambda$1.317 to be FWHM $\sim$~320 km~s$^{-1}$, 
which is practically the same as the narrow component of the recombination lines (see Table \ref{fluxes}).
For NLSy1 the widths of the O I and Fe II lines do not necessarily have to be comparable to
the broad component of the recombination lines. For example, in the case of Mrk~335,  
the Pa$\beta$ FWHM = 2050 km~s$^{-1}$, while the O I$\lambda$1.128 FWHM = 1100 km~s$^{-1}$ \citep{Rodriguez02b}.
More dramatic is the case of Ark~564, another typical NLSy1, for which the Pa$\beta$ 
FWHM = 1800 km~s$^{-1}$, while the O I$\lambda$1.128 FWHM = 600 km~s$^{-1}$ \citep{Rodriguez02b}.

\begin{figure}
\centering
\includegraphics[width=9cm]{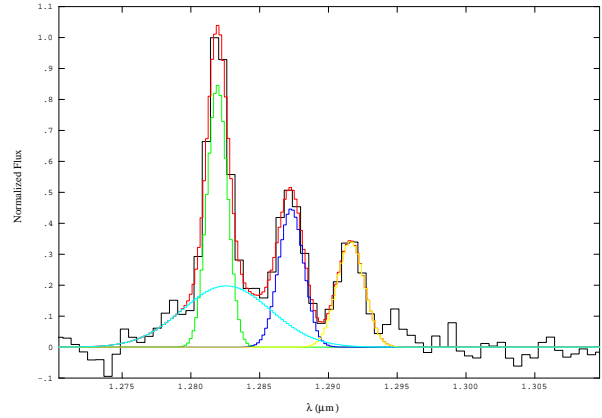}
\caption{\footnotesize{Detailed view of the nuclear spectrum of Mrk~573 around the
Pa$\beta$ line. The narrow and intermediante components are
represented, together with the gaussian profiles of the two unidentified lines, at 1.287 and 
1.292 \micron, respectively. }
\label{pab}}
\end{figure}

A relatively broad component of the Pa$\beta$ profile with FWHM $\sim$~1700 km~s$^{-1}$ 
is detected, in addition to the
narrow component (see Fig.~\ref{pab} and Table~\ref{fluxes}). This line width is
smaller than the reported by \citet{Nagao04} for the optical hydrogen recombination lines 
in polarized light. 
The  Pa$\beta$  broad component contains about 50\% of the total line flux, which is very 
similar to the fraction found by \citet{Nagar02} for the same line in NGC~5506. 


Another interesting feature detected in our spectrum is seen at 0.962 \micron,
very close to the prominent [S~III]$\lambda$0.953 line. The former is not detected in previous near-infrared
spectra of Mrk~573 \citep{Veilleux97,Riffel06} and remains unidentified in the literature. The most likely
possible identifications of this line are O I (3D$_0$-3D) or He I (1P$_0$-1D)\footnote{Computed with the 
line identification software package EMILI \citep{Sharpee03}, that employs an updated version 
of the v2.04 Atomic Line List.}. 
We can not discard either of the two posibilities, since we detect other O I lines in our near-infrared spectrum, 
and also, in the optical, He~I$\lambda$0.668 (1Po-1D) has been detected \citep{Erkens97}.
It is noticeable the FWHM of this line ($\sim$ 700 km~s$^{-1}$), that is the broadest of our spectrum (see
Table~\ref{fluxes}), which could be due to line blending.
There are another two lines very close to Pa$\beta$ (see Figs.~\ref{spectrum} and \ref{pab}), 
at 1.287 and 1.292 \micron. Tentatively, they are Ne II (2D-2P$_0$) and 
Ne I\footnote{Ne I ($2s^{2}2p^{5}(^{2}P^{o}_{1/2})3p~^{2}[3/2]_{2} - 2s^{2}2p^{5}(^{2}P^{o}_{3/2})4s~^{2}[3/2]^{o}_{2}$)}, 
respectively, according with the NIST
Atomic Spectra Database\footnote{http://physics.nist.gov/PhysRefData/ASD/lines$_{-}$form.html}.
The non-detection in previous near-infrared spectra of all the features mentioned in this section could be due 
to the higher signal-to-noise of our data, or to a more complex scenario that would allow time variations in the
spectrum of Mrk~573.

\section{Discussion}

\begin{figure}
\centering
\includegraphics[width=5cm,height=8cm,angle=90]{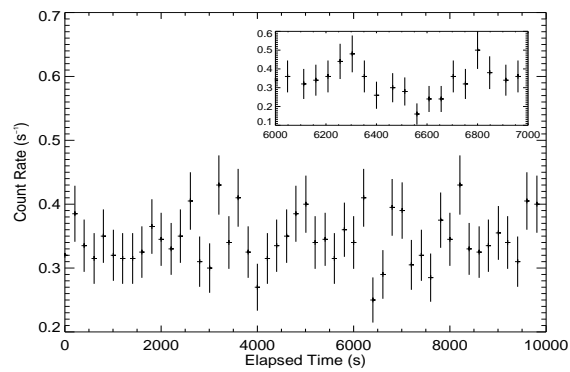}
\caption{\footnotesize{The 2-12 keV EPIC-pn light curve of Mrk~573 spanning the full length 
of the observations with bins of 200 s. The inset represent one of the extreme events with smaller
bins of 50~s.}
\label{xray}}
\end{figure}

The detection in our near-infrared spectra of the permitted Fe II and O I lines, together with the 
relatively broad component ($\sim$ 1700 km~s$^{-1}$) of Pa$\beta$ argues against the Seyfert 2 nature 
of Mrk~573 and fits perfectly in the classical definition of NLSy1 galaxies. 
This is also confirmed by the presence of the optical Fe II multiplet in polarized light \citep{Nagao04}. 
The X-ray properties of Mrk~573 are also typical of 
NLSy1 galaxies: a soft X-ray excess and a steep photon spectral index ($\Gamma$=2.7$\pm$0.4) 
\citep{Guainazzi05}. We have searched for X-ray variability in the publicly available XMM-Newton data 
of this galaxy\footnote{ The data (taken on Jan 2004) have been retrieved
from the XMM-Newton archive and processed using the task "evselect" from the package SAS v7.0.0.}. 
The light curve shows variations of the count rates by a factor of $\sim 2$ within 300~s
(see Fig.~\ref{xray}), which is similar to the variability of NLSy1s 
\citep{Leighly99,Véron01,Dewangan05}. 

The optical spectra also reveals broad components of the H$\alpha$ and H$\beta$
profiles of $\sim$~3000 km~s$^{-1}$ in polarized light \citep{Nagao04}. These line widths 
are, in principle, more typical of Sy1 than of NLSy1 galaxies. Nevertheless, very
recently, \citet{Mullaney08} found that in order to reproduce the H$\alpha$ and H$\beta$ profiles 
of a sample of well-known NLSy1 galaxies they needed three components of different widths (the broadest ones
of FWHM $\gtrsim$ 3000 km~s$^{-1}$). 
These findings indicate that the limit of FWHM~$<$ 2000~km/s imposed in the definition of NLSy1
\citep{Pogge00} is not realistic. 
Indeed, Mrk~573 shows examples of the three kinematical components in its optical and near-infrared spectra.  
Thus, there must exist an intermediate high-density region, with intermediate properties between the 
Narrow-Line Region (NLR) and the BLR, that is producing
the observed O~I and Fe~II lines, together with the relatively broad component of Pa$\beta$. 
This intermediate region must be
dense enough to produce lines which are broader than the narrow lines, but not as broad as those 
detected in polarized light.


In "classical" NLSy1 galaxies, both the O I and Fe II permitted lines
are normally more prominent \citep{Riffel06}, 
although not necessarily broader \citep{Grandi80} than in Mrk~573. 
Moreover, their optical counterparts are also detected \citep{Collin00}, which is not the case for 
this galaxy. These lines must then be produced in the previously mentioned intermediate 
high-density region, and must be affected by a moderate amount of extinction. 
More internally, in the BLR, the
$\sim$~3000 km~s$^{-1}$ components observed in polarized light \citep{Nagao04} should be produced, 	
but it is difficult to see them directly, due to the presence of an obscuring structure. 
A similar scenario occurs in the NLSy1 Mrk~618, in which  
the H$\alpha$ profile  shows three components: a
narrow one of FWHM $\sim$~200 km~s$^{-1}$, an intermediate of $\sim$~1700 km~s$^{-1}$, 
and a broad component of $\sim$~3200 km~s$^{-1}$ \citep{Mullaney08}.
For Mrk~573, we would be seeing the narrow and intermediate components, while the
broadest one would be obscured. 
The existence of an obscuring structure is also in agreement with the X-ray spectra being
Compton thick based on the large EW of the K$_\alpha$ fluorescent iron line \citep{Guainazzi05}. 
Indeed, from high-spatial resolution color maps, \citet{Quillen99} reported the presence
of a bar of length 2\arcsec, together with leading dust lanes and possibly a dusty disk perpendicular to the jet.
\citet{Guainazzi05} claimed that the presence of dust lanes is correlated with X-ray obscuration. 

From all the above considerations, we can re-classify the nucleus of Mrk~573 as a hidden NLSy1, 
a new member of the 
class proposed by \citet{Nagar02} and \citet{Dewangan05}. It shares the properties of the class: 
X-ray variability within similar timescales, steep photon spectral index, high intrinsic 
absorption column density,  and prominent iron K$_\alpha$ line.  
It is also worth to mention that in the infrared spectrum  the  
Pa$\beta$ profile  is very similar to that  of NGC~5506, 
presenting in both cases a broad pedestal (FWHM$\sim 1800$~km~s$^{-1}$).




A possible scenario that would explain the observed properties of Mrk~573 would be 
the existence of a clumpy dusty torus \citep{Nenkova02,Honig06,Honig07} that at the epoch of the 
observations was optically thinner, revealing high-density regions capable of producing the O I and Fe II lines,
together with the intermediate component of Pa$\beta$. 
The clumply torus model would allow mutations between Type 1 and Type 2 objects, 
as for NGC~7582 \citep{Aretxaga99}, which is one of the three objects classified 
by \citet{Dewangan05} as an obscured NLSy1. 
This scenario would also explain the detection of the permitted O I and Fe II lines, the relatively 
broad component of Pa$\beta$, and the unidentified 0.962, 1.287, and 1.292 \micron~features, 
that were not reported in the \citet{Riffel06} near-infrared spectrum, obtained in October 2003.

A more complex revision of this scenario was discussed by 
\citet{Elitzur06}, in which the clumpy torus would be formed as an outflow or
hydromagnetic disk wind. They proposed that the torus dissapears when the bolometric luminosity goes below 
$10^{42}$~erg~s$^{-1}$, since the accretion onto the central black hole cannot then sustain the necessary cloud outflow
rate. They also predict that for even lower luminosities the BLR would also dissapear, 
due to the suppression of the cloud outflow.
Mrk~573 could be near the lower luminosity limit required for the maintenance of the cloud outflow, 
leading to the eventual demise of the obscuring structure.

\acknowledgments

This work was partially funded by PN AYA2007-67965-C03-01, PN AYA2004-03136, and PN AYA2005-04149.

John Beckman, Andr\'{e}s Asensio Ramos, C\'{e}sar Esteban, and Bego\~{n}a Garc\'\i a-Lorenzo are 
acknowledged for their valuable help. 

The William Herschel Telescope is operated on the island of La Palma by the Isaac Newton Group in the
Spanish Observatorio del Roque de los Muchachos of the Instituto de Astrof\'\i sica de Canarias.

The authors acknowledge the data analysis facilities provided by the Starlink Project, which is run by CCLRC
on behalf of PPARC.

Based on observations obtained with XMM-Newton, an ESA science mission with instruments and 
contributions directly funded by ESA Member States and NASA.

\clearpage

\begin{table}[ !ht ]
\centering
\footnotesize
\begin{tabular}{lccc}
\hline
\hline
Line Nucleus &\multicolumn{1}{c}{$\lambda$}&\multicolumn{1}{c}{Flux}&\multicolumn{1}{c}{FWHM} \\
& (\micron) & 10$^{-15}$ ergs~cm$^{-2}$~s$^{-1}$ & km~s$^{-1}$ \\
\hline
$[S III]$	& 0.907 & 116 	$\pm$ 	3   &	 470 $\pm$ 20   \\
$[S III]$	& 0.953 & 279   $\pm$   3   &	 378 $\pm$ 8    \\
O I + He I ?    & 0.962 & 26.5  $\pm$   1.7 &	 680 $\pm$ 50   \\ 
$[C I]	$       & 0.982 & 2.9  $\pm$	0.6 &	 260 $\pm$ 90    \\ 
$[C I]	$       & 0.985 & 5.4  $\pm$	0.8 &	 250 $\pm$ 90   \\  
$[S VIII] $     & 0.991 & 13.6  $\pm$   0.8 &	 200 $\pm$ 40   \\ 
Pa$\delta$      & 1.005 & 4.4  $\pm$    0.5 &    ...	 	\\
He II	        & 1.012 & 21.3  $\pm$   0.8 &	 510 $\pm$ 30	\\ 
$[S II]$	& 1.029 & 5.6   $\pm$   0.7 &	 210 $\pm$ 80	\\
$[S II]$        & 1.032 & 9.4   $\pm$   0.9 &	 670 $\pm$ 90	\\
He I	        & 1.083 & 119    $\pm$  2   &    370 $\pm$ 14	 \\
Pa$\gamma$      & 1.094 & 16.6   $\pm$  1.6 &    430 $\pm$ 60	 \\
O I 		& 1.128 & 2.6   $\pm$   0.9 &	...	       \\  
$[P II]$        & 1.147 & 3.3   $\pm$   0.7 &	340 $\pm$ 140	\\  
$[P II]$	& 1.188 & 3.4   $\pm$   0.7 &	...	       \\  
$[S IX]$	& 1.252 & 8.5   $\pm$   0.7 &	200 $\pm$ 40	\\ 
$[Fe II]$	& 1.257 & 17    $\pm$   1   &	620 $\pm$ 50   \\ 
Pa$\beta$ narrow& 1.282 & 15.7   $\pm$  1.3 &	...	       \\
Pa$\beta$ broad	& 1.282 & 16     $\pm$  3   &	1700 $\pm$ 400  \\
Ne II ?      	& 1.287 & 10.3   $\pm$  1.7 &	270  $\pm$ 50  \\ 
Ne I ?     	& 1.292 & 8.6   $\pm$   0.9 &   350  $\pm$ 60	\\
O I 		& 1.317 & 8     $\pm$   2   &   320  $\pm$ 100 \\
$[Fe II]$ 	& 1.321 & 13     $\pm$  3   &   480  $\pm$ 160 \\
\hline	   						     																																																									       
\end{tabular}	
\caption{\footnotesize{Emission lines detected, central wavelength (rest frame), line fluxes, 
and FWHM corrected for instrumental broadening. Lines whose FWHM values are not shown are narrower than
the instrumental profile.}
\label{fluxes}}
\end{table}

\clearpage

\end{document}